\documentclass[aps,prd,groupedaddress,amssymb,twocolumn,eqsecnum,showpacs,epsfig,nofootinbib]{revtex4}

\usepackage{graphicx}
\usepackage{bm}
\usepackage{dcolumn}
\usepackage{amsmath}

\usepackage{amsmath}
\usepackage{amssymb}

\numberwithin{equation}{section}
\usepackage{epsfig}

\begin{document}
\newcommand{\newc}{\newcommand}

\newc{\be}{\begin{equation}}
\newc{\ee}{\end{equation}}
\newc{\ba}{\begin{eqnarray}}
\newc{\ea}{\end{eqnarray}}
\newc{\bea}{\begin{eqnarray*}}
\newc{\eea}{\end{eqnarray*}}
\newc{\D}{\partial}
\newc{\ie}{{\it i.e.} }
\newc{\eg}{{\it e.g.} }
\newc{\etc}{{\it etc.} }
{\newc{\etal}{{\it et al.}}
\newc{\lcdm}{$\Lambda$CDM}
\newcommand{\nn}{\nonumber}
\newc{\ra}{\rightarrow}
\newc{\lra}{\leftrightarrow}
\newc{\lsim}{\buildrel{<}\over{\sim}}
\newc{\gsim}{\buildrel{>}\over{\sim}}
\newcommand{\mincir}{\raise
-3.truept\hbox{\rlap{\hbox{$\sim$}}\raise4.truept\hbox{$<$}\ }}
\newcommand{\magcir}{\raise
-3.truept\hbox{\rlap{\hbox{$\sim$}}\raise4.truept\hbox{$>$}\ }}

\title{Dynamics and Constraints of the Massive Gravitons Dark Matter Flat Cosmologies}

\author{S. Basilakos}\email{svasil@academyofathens.gr}
\affiliation{Academy of Athens, Research Center for Astronomy and
Applied Mathematics,
 Soranou Efesiou 4, 11527, Athens, Greece}

\author{M. Plionis}\email{mplionis@astro.noa.gr}
\affiliation{Institute of Astronomy \& Astrophysics, Nationals
Observatory of Athens, Thessio 11810, Athens, Greece, and
\\Instituto Nacional de Astrof\'isica, \'Optica y Electr\'onica, 72000 Puebla, Mexico}

\author{M. E. S. Alves}\email{alvesmes@unifei.edu.br}

\affiliation{Instituto de Ci\^encias Exatas, Universidade Federal de Itajub\'a \\
Av. BPS, 1303, 37500-903, Itajub\'a, MG, Brazil}

\author{J. A. S. Lima}\email{limajas@astro.iag.usp.br}

\affiliation{Departamento de Astronomia (IAGUSP),  Universidade de S\~ao Paulo\\
Rua do Mat\~ao, 1226, 05508-900, S. Paulo, Brazil}

\begin{abstract}
We discuss the dynamics of the universe within the framework of
Massive Graviton Dark Matter scenario (MGCDM) in which
gravitons are geometrically treated as massive particles.
In this modified gravity theory,  the main effect of the gravitons
is to alter the  density evolution of the cold dark matter component
in such a way that the Universe evolves to an accelerating expanding
regime, as presently observed. Tight constraints on the main
cosmological parameters of the MGCDM model are derived by performing a
joint likelihood analysis involving the recent supernovae type Ia
data, the Cosmic Microwave Background (CMB) shift parameter and the
Baryonic Acoustic Oscillations (BAOs) as traced by the Sloan Digital
Sky Survey (SDSS) red luminous galaxies. The linear evolution of small
density fluctuations  is also analysed in detail.  It is found that the
growth factor of the MGCDM model is slightly different ($\sim1-4\%$)
from the one provided  by the conventional flat $\Lambda$CDM cosmology.
The  growth rate of clustering  predicted by  MGCDM and $\Lambda$CDM models
are confronted to the observations  and the corresponding best fit
values of the growth index ($\gamma$) are also  determined. By using
the expectations of realistic future X-ray and Sunyaev-Zeldovich
cluster surveys we derive the dark-matter halo mass function and the
corresponding redshift distribution of cluster-size halos for the
MGCDM model.  Finally, we also show that the Hubble flow differences
between the MGCDM and the $\Lambda$CDM models provide a halo redshift
distribution  departing significantly from the ones predicted by other
DE models. These results suggest that the MGCDM model can
observationally be distinguished from  $\Lambda$CDM and also from a large
number of dark energy models recently proposed in the literature.

\end{abstract}
\pacs{98.80.-k, 95.35.+d, 95.36.+x}
\keywords{Cosmology; dark energy; large scale structure of the
Universe}
\maketitle

\section{Introduction}
The high-quality cosmological observational data (e.g. supernovae type
Ia, CMB, galaxy clustering, etc), accumulated during the last two decades,
have enabled cosmologists to gain substantial confidence that modern
cosmology is capable of quantitatively reproducing the details of
many observed cosmic phenomena, including the late time
accelerating stage of the Universe. Studies by many authors
have converged to a cosmic expansion history involving a spatially
flat geometry and a cosmic dark sector formed by cold dark matter and some
sort of dark energy, endowed with large negative pressure, in order to
explain the observed accelerating expansion of the Universe
\cite{Riess07,Spergel07,essence,Kowal08,komatsu08,Hic09,LJC09,BasPli10}.

In spite of that, the absence of a fundamental physical theory, regarding
the mechanism inducing the cosmic acceleration, have given rise to a
plethora of alternative cosmological scenarios. 
Most are based either on the existence of new fields in nature (dark
energy) or in some modification of Einstein's general relativity,
with the present accelerating stage appearing as a sort of geometric effect.

The simplest dark energy candidate corresponds to a cosmological
constant, $\Lambda$ (see \cite{reviews} for reviews).
In the standard concordance cosmological ($\Lambda$CDM) model, the
overall cosmic fluid contains baryons, cold dark matter plus a
vacuum energy. This model fits accurately the current
observational data and it therefore provides an excellent scenario
to describe the observed universe. However, it is well known that
the concordance model suffers from, among others \cite{Peri08},
two fundamental problems:

{(i) {\it Fine tuning problem} -
the fact that the observed value of the vacuum energy density
($\rho_{\Lambda}=\Lambda c^{2}/8\pi G\simeq 10^{-47}\,GeV^4$) is
more than 120 orders of magnitude below the natural value
estimated using quantum field theory \cite{Weinberg89}.

(ii) {\it Coincidence problem} - the fact
that the matter and the vacuum energy densities are of the same
order just prior to the present epoch \cite{coincidence}.}

Such problems have inspired
many authors to propose alternative dark energy candidates such as
$\Lambda(t)$ cosmologies, quintessence, $k-$essence, vector
fields, phantom dark energy, tachyons, Chaplygin gas and the list
goes on (see \cite{Ratra88,Oze87,Lambdat,Bas09c,Wetterich:1994bg,
Caldwell98,Brax:1999gp,KAM,fein02,Caldwell,Bento03,chime04,Linder2004,
Brookfield:2005td,Grande06,Boehmer:2007qa} and references
therein). Naturally, in order to establish  the evolution of the
dark energy equation of state (EoS), a realistic form of $H(a)$ is
required which should be constrained through a combination of
independent dark energy probes.

Nevertheless, there are other possibilities to explain the
present accelerating stage.  For instance, one may consider
that the dynamical effects attributed to dark energy can be mimicked
by a nonstandard gravity theory. In other words, the present
accelerating stage of the universe can also be driven only by cold
dark matter under a modification of the nature of gravity.
Such a reduction of the so-called dark
sector is naturally obtained in the so-called $f(R)$ gravity
theories \cite{FR} (see, however, \cite{Nat10}).

On the other hand,  general relativity  predicts
that gravitational waves are non-dispersive and propagate with  the
same vacuum light speed. These results lead to the common believe that
the graviton (the ``boson'' for general relativity),
must be a massless particle. However, massive gravitons are features
of some alternatives to general relativity as the one proposed by
Visser \cite{vis1998}. Such theories have motivated many experiments
and observations in order to detect a possible dispersive behavior
due to a non-zero graviton  mass (see \cite {HL2010}  and Refs. there in).

More recently, it was shown that the massive graviton approach proposed  by Visser can be used
to build realistic cosmological models that can then be tested
against the available cosmological data \cite{Alves10}. One of the main advantages  of such
massive graviton cosmology  is the fact that it contains the same number of free
parameters as the concordance $\Lambda$CDM model, and, therefore,  it does not
require the introduction of any extra fields in its dynamics.  In this way, since  the astronomical community is planning a
variety of large observational projects intended to test and constrain the standard $\Lambda$CDM
concordance model, as well as many of the proposed alternative models, it is timely and
important to identify and explore a variety of physical mechanisms (or substances) which
could also be responsible for the late-time acceleration of the Universe.

In what follows we focus our attention to a cosmological model
within Visser's massive graviton theory.
In particular we discuss how to differentiate the massive graviton
model from the concordance $\Lambda$CDM model. Initially, a joint
statistical analysis, involving the latest observational data
(SNIa, CMB shift parameter and BAO) is implemented. Secondly, we attempt to discriminate the MGCDM and
$\Lambda$CDM models by computing the halo mass function and the
corresponding redshift distribution of the cluster-size halos.
Finally, by using future X-ray and SZ surveys
we show that the evolution of the cluster abundances
is a potential discriminator between the MGCDM and $\Lambda$CDM
models. We would like to stress here that
the abundance of collapsed structures, as a
function of mass and redshift, is a key statistical test for studies
of the matter distribution in the universe, and, more importantly, it can
be accessed through observations \cite{Evra}. Indeed, the mass
function of galaxy clusters has been measured based on X-ray surveys
\cite{Borg01, Reip02, Vik09}, via weak and strong lensing studies
\cite{Bat98, Dahle06, Corl09}, using optical surveys, like the SDSS
\cite{Bah03, Wen10}, as well as, through Sunayev-Zeldovich (SZ)
effect \cite{Taub05}. In the last decade many authors have been
involved in this kind of studies and have found that the abundance
of the collapsed structures is affected by the presence of a dark energy component
\cite{Wein03,Liberato,manera,Abramo07,Fran08,Sch09,Mort09,Rap10,Pace10,Alam10,Khed10,BPL10,Lomb10}.

The paper is planned as follows.
{The basic elements of
 Visser's theory are presented in section \ref{sec:two}, where we
also introduce the cosmological equations for a flat Friedmann-Lemaitre-Robertson-Walker (FLRW) geometry with massive gravitons.}
In section \ref{sec:three}, a joint statistical analysis based on SNe Ia, CMB and BAO
is used to constraint the massive graviton cosmological model
free parameter.
The linear growth factor of matter perturbations
is discussed in section \ref{sec:four}, while in  \ref{sec:five}, we discuss
and compare the corresponding theoretical predictions regarding the evolution of
the cluster abundances. Finally, the main conclusions are
summarized in  section \ref{sec:six}.

\section{\label{sec:two}Massive Gravitons Cold Dark Matter
(MGCDM) Cosmology: Basic Equations}

 In this section we briefly present the main points of Visser's
massive gravity approach \cite{vis1998}.
The full action is given by (in what follows $\hbar = c=1$)
\begin{eqnarray}
\label{fullaction}
S=\int d^4x\left[\sqrt{-g}\frac{R(g)}{16\pi G}
+ {\cal{L}}_{mass_g}(g,g_0) +{\cal{L}}_{matter}(g)\right]
\end{eqnarray}
where besides the Einstein-Hilbert Lagrangian and the Lagrangian
of the matter fields, we have the bi-metric Lagrangian:
\begin{eqnarray}
{\cal{L}}_{mass}(g,g_0) = \frac{1}{2}{m_g}^2
\sqrt{-g_0}\bigg\{ ( g_0^{-1})^{\mu\nu}
( g-g_0)_{\mu\sigma}( g_0^{-1})^{\sigma\rho}
\\ \nonumber
 \times ( g-g_0)_{\rho\nu}-\frac{1}{2}
\left[( g_0^{-1})^{\mu\nu}( g-g_0)_{\mu\nu}\right]^2\bigg\},
\end{eqnarray}
where $m_g$ is the graviton mass and
$(g_0)_{\mu\nu}$ is a general flat metric.

The field equations, which are obtained by variation of
(\ref{fullaction}), can be written as:
\begin{equation}\label{field-equations}
G^{\mu\nu} -\frac{1}{2}{m_g}^2 M^{\mu\nu} = -{8\pi G}  T^{\mu\nu},
\end{equation}
where $G^{\mu\nu}$ is the Einstein tensor, $T^{\mu\nu}$ is the
energy-momentum tensor for perfect fluid, and the contribution of
the massive tensor to the field equations reads:
\begin{eqnarray}\label{massive tensor}
M^{\mu\nu} =  (g_0^{-1})^{\mu\sigma}\bigg[ (g-g_0)_{\sigma\rho}
- \frac{1}{2}(g_0)_{\sigma\rho}(g_0^{-1})^{\alpha\beta}\\
\nonumber
\times(g-g_0)_{\alpha\beta} \bigg](g_0^{-1})^{\rho\nu}    .
\end{eqnarray}

Note that if one takes the limit $m_g\rightarrow 0$ the standard
Einstein field equations are recovered.

Thus, from the construction of the Visser's theory, it can be classified as a 
bimetric theory of gravitation. This kind of theory was first studied by N. Rosen \cite{Rosen1973}. 
In the Rosen's concept the metric $g_{\mu\nu}$ describes the geometry of the spacetime in the same 
way as in the context of the general relativity theory, and the second metric $(g_0)_{\mu\nu}$ 
(that Rosen denoted by $\gamma_{\mu\nu}$) refers to the flat spacetime and describes the inertial 
forces. It is worth to mention that Rosen has shown that a bimetric theory satisfies the 
covariance and the equivalence principles, a fact that was also pointed out by Visser (for more 
discussion see \cite{Rham10}). 
In this way, in order to follow the Rosen's approach we have constrained the background 
metric to respect the Riemann-flat condition, that is, $R^\lambda_{\mu\nu\kappa}(g_0) \equiv 0$ 
in such a way that we have no ambiguity on the choice of $(g_0)_{\mu\nu}$, it will always be 
chosen to be a flat metric, depending only on the particular coordinates we are dealing, of course. 

Regarding the energy-momentum conservation we will follow the same
approach of Refs. \cite{Narlikar1984,deAraujo2007}.
 Since the Einstein tensor satisfies the Bianchi identities
$\nabla_\nu G^{\mu\nu} = 0$, the energy conservation law is expressed as:
 \begin{equation}\label{conservation}
   \nabla_\nu T^{\mu\nu} = \frac{{m_g}^2}{16\pi G} \nabla_\nu M^{\mu\nu}.
 \end{equation}

In the above framework, the global dynamics of a flat MGCDM cosmology
is driven by the following equations\footnote{In the present article we restrict our analysis 
to the flat cosmologies in order to compare our results with those of the flat $\Lambda$CDM 
model that is the most accepted cosmological model as shown, e.g., by the WMAP7 data \cite{komatsu08}. 
A generalization of the model for a non spatially flat cosmology will appear in a forthcoming article.}:

 \begin{equation}\label{eqfried1}
 {8\pi G \rho}  = 3\left( \frac{\dot{a}}{a}\right)^2 + \frac{3}{4}{m_g}^2(a^2 - 1),
\end{equation}
 \begin{equation}\label{eqfried2}
{8\pi G p} =  -2\frac{\ddot{a}}{a} - \left( \frac{\dot{a}}{a}\right)^2 - \frac{1}{4}{m_g}^2a^2(a^2-1),
 \end{equation}
where $\rho$ is the energy density, $p$ is the
pressure and  $a(t)$ is the scale factor.

From Eq. (\ref{conservation}) we get the evolution equation for
the energy density, namely:
 \begin{equation}
   \dot{\rho} + 3 H \left[  (\rho + p) + \frac{{m_g}^2}{32\pi G} (a^4 - 6a^2 + 3) \right] = 0,
 \end{equation}
where $H = \dot{a}/a$. By integrating the above
equation for a matter dominated universe ($p= 0$)
one obtains:
\begin{equation}\label{rho-m-new}
\rho(a) = \frac{\rho_0}{a^3}- \frac{3{m_g}^2}{32\pi G}
\left( \frac{a^4}{7} - \frac{6a^2}{5} + 1 \right) \;,
\end{equation}
where $\rho_0$ is the present value of the energy density.
As expected, in the limiting case $m_g \rightarrow 0$  all
the standard FLRW expressions are recovered.

Now, inserting (\ref{rho-m-new}) in the modified Friedmann
equation (\ref{eqfried1}) we obtain the normalized Hubble parameter:
 \begin{equation}\label{parHubMass}
E^{2}(a)=\frac{H^{2}(a)}{H_{0}^{2}}= \Omega_{m}a^{-3}+\delta H^{2},
\end{equation}
with
 \begin{equation}\label{parHubMass1}
\delta H^{2}= \frac{1}{2}\Omega_{g} \left( 7a^2 -5a^4 \right),
 \end{equation}
where $H_0$ is the Hubble constant, $\Omega_m$ is the matter
density parameter (for baryons and dark matter
$\Omega_i=\rho_{i0}/\rho_{c0}$, where $\rho_{c0}=3{H_0}^{2}/8\pi G$
is the critical density parameter), and $\Omega_{g}=\frac{1}{70}
(\frac{m_g}{H_0})^{2}$ is
the present contribution of the massive gravitons.
It should be stressed that the last term of the above
normalized Hubble function (\ref{parHubMass}) encodes
the correction to the standard FLRW expression.

In general, using the FLRW equations, one can express
the effective dark energy EoS parameter in terms of the
normalized Hubble parameter \cite{Saini00}
\begin{equation}
\label{eos22}
w_{\rm DE}(a)=\frac{-1-\frac{2}{3}a\frac{{d\rm ln}E}{da}}
{1-\Omega_{m}a^{-3}E^{-2}(a)}.
\end{equation}
After some simple algebra, it is also readily seen that
the effective (``geometrical'' in our case) dark energy EoS parameter is
given by (see \cite{Linjen03, Linder2004}):
\begin{equation}
\label{eos223}
w_{\rm DE}(a)=-1-\frac{1}{3}\;\frac{d{\rm ln}\delta
H^{2}}{d{\rm ln}a}.
\end{equation}

In our case, inserting Eq. (\ref{parHubMass1}) into Eq.
(\ref{eos223}) it is straightforward to obtain a simple analytical
expression for the geometrical dark energy EoS parameter:
\begin{equation}\label{eos221}
w_{\rm DE}(a) = -  1 - \frac{2}{3} \left(\frac{7 - 10a^2}{7 - 5a^2}\right).
\end{equation}

It thus follows that  in the cosmological context, the modified
gravity theory as proposed by Visser can be
treated as an additional effective fluid with EoS parameter defined by (\ref{eos221}). Note also that the current Hubble function has
only two free parameters ($H_0$ and $\Omega_{m}$),
exactly the same number of free parameters as the conventional flat $\Lambda$CDM
model. Naturally, the value of $H_0$ is not predicted by any of the
models and it is set to its observational value of $H_0=70.4$ km
s$^{-1}$ Mpc$^{-1}$ \cite{komatsu08,freedman}.

\section{\label{sec:three}Likelihood Analysis}
Let us now discuss the statistical treatment of the observational data used
to constrain the MGCDM model presented in the previous section.

To begin with, we consider the {\em Constitution} supernovae Ia set of Hicken
et al. \cite{Hic09}, but in order to avoid possible problems
related to the local bulk flow, we use a subset of this sample
containing 366 SNe Ia all with redshifts $z>0.02$. The likelihood
estimator is determined by a $\chi^{2}_{\rm SNIa}$ statistics:
\begin{equation}
\label{chi22} \chi^{2}_{\rm SNIa}(\Omega_{m})=\sum_{i=1}^{366} \left[
\frac{ {\cal \mu}^{\rm th} (a_{i},{\Omega_{m}})-{\cal \mu}^{\rm
obs}(a_{i}) } {\sigma_{i}} \right]^{2},
\end{equation}
where $a_{i}=(1+z_{i})^{-1}$ is the scale factor of the Universe at
the observed redshift $z_{i}$, ${\cal \mu}$ is the distance modulus
${\cal \mu}=m-M=5{\rm log}d_{L}+25$ and $d_{L}$ is the
luminosity distance\footnote{Since
only the relative distances of the SNIa are accurate and not their
absolute local calibration, we always marginalize with respect to the
internally derived Hubble constant.}, $ d_{L}(a,\Omega_{m})=c{a}^{-1} \int_{a}^{1}
\frac{{\rm d}y}{y^{2}H(y)}$.
Now, from the likelihood analysis we find that
$\Omega_{m}=0.266\pm 0.016$ with
$\chi_{tot}^{2}(\Omega_{m})/dof\simeq 446.5/365$.

In addition to the SNe Ia
data, we also consider the BAO scale produced in the last
scattering surface by the competition between the pressure of the
coupled baryon-photon fluid and gravity. The resulting acoustic
waves leave (in the course of the evolution) an overdensity
signature at certain length scales of the matter distribution.
Evidence of this excess was recently found in the clustering
properties of SDSS galaxies (see \cite{Eis05,Perc10,Kazin10})
and it provides a suitable ``standard ruler'' for constraining dark energy
models. In this work we use
the measurement derived by Eisenstein et al. \cite{Eis05}. In particular,
we utilize the following estimator $A(\Omega_{m})=
\frac{\sqrt{\Omega_{m}}}{[z^{2}_{s}E(a_{s})]^{1/3}}
\left[\int_{a_{s}}^{1} \frac{da}{a^{2}E(a)} \right]^{2/3}$, measured
from the SDSS data to be $A=0.469\pm 0.017$, where $z_{s}=0.35$ [or
$a_{s}=(1+z_{s})^{-1}\simeq 0.75$]. Therefore, the corresponding
$\chi^{2}_{\rm BAO}$ function can be written as:
\begin{equation}
\chi^{2}_{\rm BAO}(\Omega_{m})=\frac{[A(\Omega_{m})-0.469]^{2}}{0.017^{2}}\;.
\end{equation}
The likelihood function peaks at $\Omega_{m}=0.306^{+0.026}_{-0.025}$.

Finally, a very interesting geometrical probe of
dark energy is provided by the angular scale of the sound horizon at
the last scattering surface. It is  encoded in the location
of the first peak of the angular (CMB) power spectrum
\cite{Bond:1997wr,Nesseris:2006er}, and may be defined by the
quantity ${\cal R}=\sqrt{\Omega_{m}}\int_{a_{ls}}^1 \frac{da}{a^2 E(a)}$.
The shift parameter measured from the WMAP 7-years data
\cite{komatsu08} is ${\cal R}=1.726\pm 0.019$ at $z_{ls}=1091.36$ [or
$a_{ls}=(1+z_{ls})^{-1}\simeq 9.154\times 10^{-4}$]. In this case,
the $\chi^{2}_{\rm cmb}$ function reads
\begin{equation}
\chi^{2}_{\rm cmb}(\Omega_{m})=\frac{[{\cal R}(\Omega_{m})-1.726]^{2}}{0.018^{2}}.
\end{equation}
It should be stressed that for CMB shift parameter, the
contribution of the radiative component, ($\Omega_{R} a^{-4}$,
where $\Omega_{R}\simeq 4.174\times 10^{-5}h^{-2}$) needs also to
be considered \cite{komatsu08}. Note also that the measured CMB
shift parameter is somewhat model dependent but such
details of the models were not included in our analysis. For
example, such is the case when massive neutrinos are included.
The robustness of the shift parameter
has been tested and discussed in \cite{Elgaroy07}.
In this case the best fit value is: $\Omega_{m}=0.263 \pm 0.03$.

The derived $\Omega_m$ values from each individual probe appear to be
quite different, although within their mutual 2$\sigma$ uncertainty
range. Therefore, in order to put tighter constraints on the corresponding
parameter space of any cosmological model, the above probes are combined
through a joint likelihood analysis\footnote{Likelihoods are
normalized to their maximum values. In the present analysis we
always report $1\sigma$ uncertainties on the fitted parameters. Note
also that the total number of data points used here is
$N_{tot}=368$, while the associated degrees of freedom are: {\em
  dof}$= 367$. Note that we sample $\Omega_{m} \in [0.1,1]$ in steps of
0.001.}, given by the product of the individual likelihoods
according to: ${\cal L}_{tot}(\Omega_{m})= {\cal L}_{\rm SNIa}\times
{\cal L}_{\rm BAO} \times {\cal L}_{\rm cmb}$, which translates in the
joint $\chi^2$ function in an addition:
$\chi^{2}_{tot}(\Omega_{m})=\chi^{2}_{\rm SNIa}+\chi^{2}_{\rm
BAO}+\chi^{2}_{\rm cmb}$.

Now, by applying our joint statistical procedure for both cosmologies,
we obtain the following best fit parameters:
\begin{itemize}
\item MGCDM model: $\Omega_{m}=0.276\pm 0.012$
with $\chi_{tot}^{2}(\Omega_{m})/dof \simeq
448.5/367$.  Such results should be compared to those found
by Alves et al. \cite{Alves10}, namely: $\Omega_{m}=0.273\pm 0.015$
with a $\chi_{tot}^{2}(\Omega_{m})/dof \simeq
565.06/558$. This difference must be probably attributed
to the use of the {\em Union2} supernovae sample \cite{Union2} by the
latter authors.
\item $\Lambda$CDM model: $\Omega_{m}=0.280\pm 0.010$ with
$\chi_{tot}^{2}(\Omega_{m})/dof\simeq 439.5/367$,
which is in good agreement with recent studies
\cite{Riess07,Spergel07,essence,Kowal08,komatsu08,Hic09,LJC09,BasPli10}.
\end{itemize}

\begin{figure}[ht]
\mbox{\epsfxsize=8.5cm \epsffile{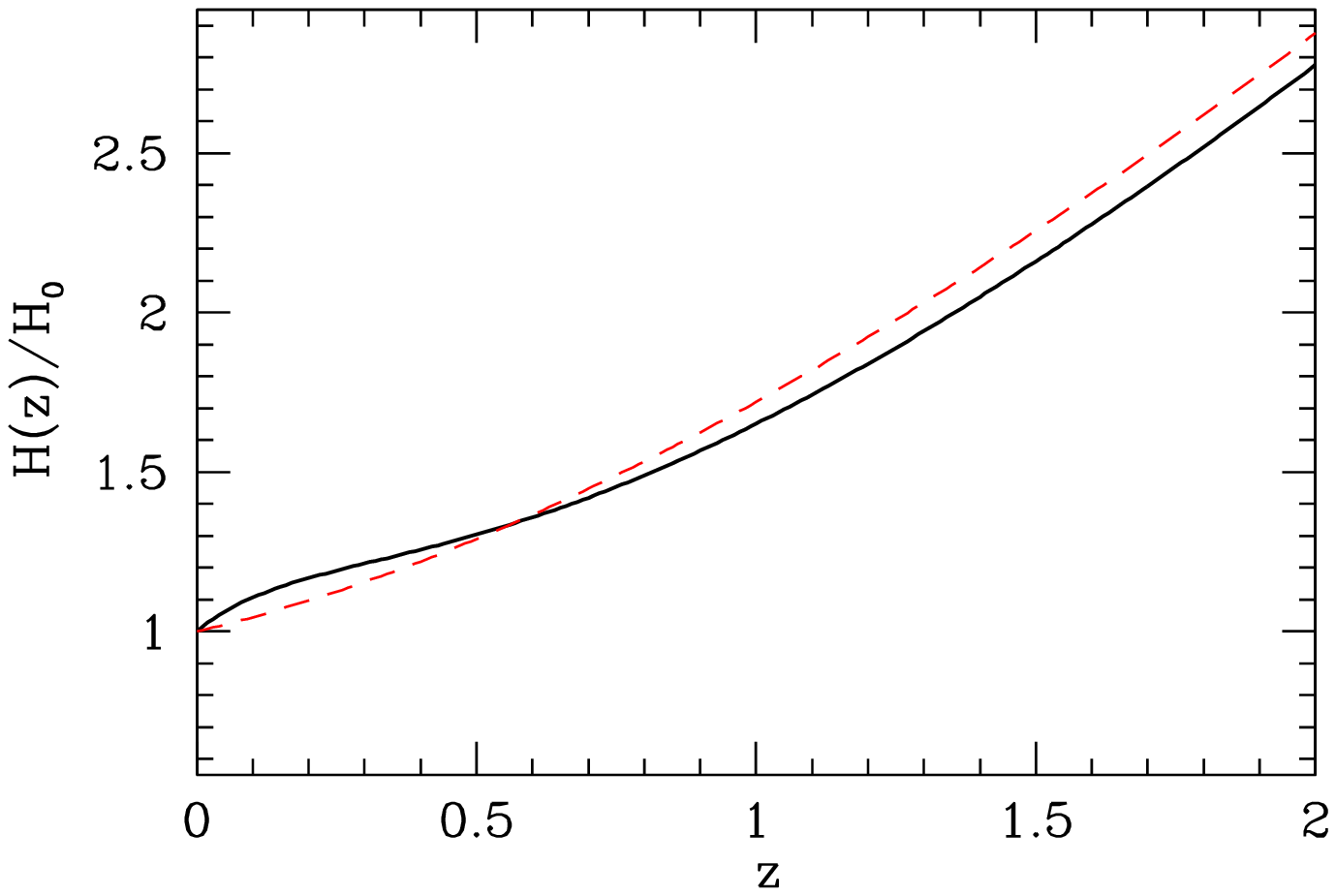}}
\caption{Normalized Hubble parameter as a function of redshift.
The solid line is the prediction of the MGCDM model.
For comparison, the dashed line corresponds to
the traditional $\Lambda$CDM model.}
\end{figure}

\begin{figure}[ht]
\mbox{\epsfxsize=8.5cm \epsffile{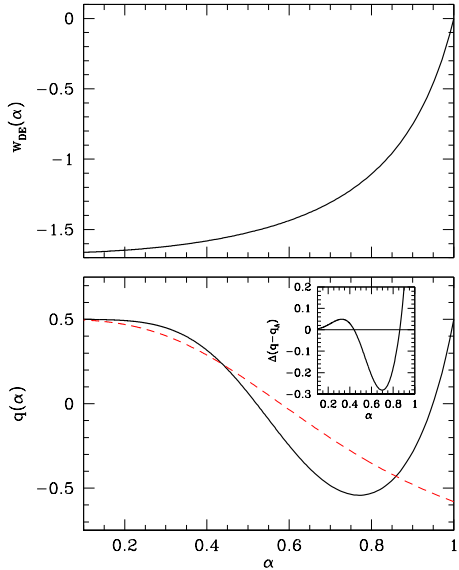}}
\caption{Expansion history. In the upper panel
we display the evolution of the dark energy effective
EoS parameter. In the lower panel we compare the
deceleration parameters of the MGCDM (solid line)
and the concordance $\Lambda$CDM (dashed line) models.
In the insert we show the relative deviation
$\Delta(q-q_{\Lambda})$ of the two deceleration parameters.}
\end{figure}

It should be mentioned here that using the BAO results
of Percival et al. \cite{Perc10}, does not change the previously
presented constraints.

\section{\label{sec:four}MGCDM versus $\Lambda$CDM cosmology}

\subsection{The cosmic expansion history}
 In Figure 1 we plot the normalized MGCDM Hubble function (solid line)
as a function of redshift, which appears quite different both in amplitude
and shape with respect to the corresponding $\Lambda$CDM model expectations
(dashed-line).

In figure 2 (upper panel), we  present the evolution of the MGCDM effective dark energy EoS
parameter. 
One can divide the evolution of the cosmic expansion history in different
phases on the basis of the varying  behavior of the
MGCDM and $\Lambda$CDM models. We will investigate such variations
in terms of the deceleration parameter, $q(a)=-(1+d{\rm ln}H/d{\rm
  ln}a)$, which is plotted in the lower panel of figure 2. In the inset plot
we display the relative deviation of the deceleration parameter,
$\Delta(q-q_{\Lambda})$, between the two cosmological models.
We can divide the cosmic expansion history in the following phases:
\begin{itemize}

\item at early enough times $a\mincir 0.1$ the deceleration
    parameters of both models are positive with $q\simeq
    q_{\Lambda}$, which means that the two cosmological models
    provide a similar expansion rate of the universe.
    Note that by taking the limit $\displaystyle
    \lim_{a \to 0} w_{DE}(a) = -5/3$ for the MGCDM model, while
    we always have $w_{DE} = -1$ for the $\Lambda$CDM model;
\item for $0.1\le a\le 0.44$ the deceleration parameters are
    both positive with $q>q_{\Lambda}$, which means that the
    cosmic expansion in the MGCDM model is more rapidly
    ``decelerating'' than in the $\Lambda$CDM case;
\item between $0.44<a<0.52$ the deceleration parameters remain
    positive but $q<q_{\Lambda}$;
\item for $0.52\le a\le 0.57$ the traditional $\Lambda$ model
    remains in the decelerated regime ($q_{\Lambda}>0$) but
    the MGCDM is starting to accelerate ($q<0$);
\item for $0.57<a\le 0.94$ the deceleration parameters are
    both negative and since $q<q_{\Lambda}$, the
    MGCDM model provides a stronger acceleration than in the
    $\Lambda$CDM model (the opposite situation holds at
    $0.85\le a\le 0.94$).
\end{itemize}
Interestingly, prior to the present epoch ($a>0.94$) the the
deceleration parameter of the MGCDM model becomes positive
and when $a = 1$ we have $w_{DE} = 0$, ie., the universe
becomes again matter dominated, implying that the late time
acceleration of the universe was a transient phase which has
already finished.

From the inset panel of figure 2 it becomes clear that the MGCDM
model reaches a maximum deviation from the $\Lambda$CDM cosmology
prior to $a \simeq 0.75$ and again at $a \simeq 1$. Finally, the
deceleration parameters at the present time are $q_{0}\simeq 0.50$
and $q_{0\Lambda}\simeq -0.58$. If we go further to the
future we find from Eq. (\ref{eos221}) that the state parameter as
well as the deceleration parameter diverges for $a = \sqrt{7/5}$.
This value sets the turning point after which the universe begins to
contract in the MGCDM model (for more details see
\cite{Alves2009}.)

\subsection{The growth factor and the rate of clustering}

It is well known that for small scales (smaller than the horizon)
the dark energy component (or ''geometrical'' dark energy)
is expected to be smooth and thus it is
fair to consider perturbations only on the matter component of the
cosmic fluid \cite{Dave02}. This assumption leads to the usual 
equation for matter perturbations
\begin{equation}
\label{fluc01}
\ddot{\delta}_m+2H\dot{\delta}_m-4\pi G_{\rm eff} \rho_{m}\delta_m=0,
\end{equation}
where the effect of ''geometrical'' dark energy 
is introduced via the expression of $G_{\rm eff}=G_{\rm eff}(t)$ [see \cite{Lue04},\cite{Tsu08}].
In the context of general relativity $G_{\rm eff}$ coincides with 
the Newton's gravitational constant.
Now, for any type of dark energy 
an efficient parametrization of the matter perturbations ($\delta_{m} \propto D$)
is based on the growth rate $f(a)\equiv d{\rm ln}D/d{\rm ln}a$
\cite{Peeb93}, which has the following functional form:
\be
\label{fzz221}
f(a)=\frac{d{\rm ln}D}{d{\rm ln}a}=\Omega^{\gamma}_{m}(a) \;\;,
\ee
where $D(a)$ is the linear growth factor,  
$\Omega_{m}(a)=\Omega_{m}a^{-3}/E^{2}(a)$ and $\gamma$ is the so
called growth index (see Refs. \cite{Linder2004,Linjen03,Wang98,Lue04,Linder2007}).
Since the growth factor of a pure matter universe (Einstein de-Sitter)
has the form $D_{\rm EdS}=a$, one has to normalize the different
cosmological models such that $D\simeq a$ at
large redshifts due to the dominance of the non-relativistic matter component.
Using the latter condition we can easily integrate Eq. (\ref{fzz221})
to derive the growth factor \cite{Linder2004}
\be
\label{Dz221}
D(a)=a {\rm e}^{\int_{0}^{a} (dx/x) [\Omega_{m}^{\gamma}(x)-1]} \;.
\ee

In the present case we are working with a modification of
Einstein's gravity instead of an extra fluid, in such a way the
usual Poisson equation for the gravitational potential is modified
due the presence of the mass term. In the simple case of the
non-relativistic limit we have a Yukawa-like potential which
accomplishes corrections to the Newtonian potential to scales of
the order of the Compton wavelength of the graviton, $\lambda =
m^{-1}_{g}$. Using this kind of potential, the classic limit for the
graviton mass obtained from solar system dynamics observations is
$m_g < 7.68 \times 10^{-55}$g \cite{Talmadge1988}, but one of the
most stringent constraints is obtained by requiring the derived
dynamical properties of a galactic disk to be consistent with
observations \cite{deAraujo2007} thereby yielding $m_g < 10^{-59}$g. Now, by
considering the best fit value obtained here for $\Omega_g$ we
have $m_g \sim 10^{-65}$g, which is nearly 6 orders of magnitude
below to the previous bound.
This value gives a Compton wavelength of the order of the horizon. The 
Compton wavelength can be seen as the physical length of graviton's perturbations which 
implies that these perturbations play some role only close to the Hubble radius
and thus they will be negligible at sub-horizon scales. In other words, 
Eqs. (\ref{fluc01}), (\ref{Dz221}) are both valid also in the MGCDM model.

\begin{table}[ht]
\caption[]{Data of the growth rate of clustering \cite{Ness08}. The
correspondence of the columns is as follows: redshift, observed
growth rate and references.} \tabcolsep 4.5pt
\begin{tabular}{ccc} \hline \hline
z& $f_{obs}$ & Refs. \\ \hline
0.15 & $0.51\pm 0.11$& \cite{Verde02,Hawk03}\\
0.35 & $0.70\pm 0.18$& \cite{Teg06} \\
0.55 & $0.75\pm 0.18$& \cite{Ross07}\\
1.40 & $0.90\pm 0.24$& \cite{daAng08}\\
3.00 & $1.46\pm 0.29$& \cite{McDon05}\\
\end{tabular}
\end{table}

Clearly in order to quantify the evolution of the growth
factor we need to know the growth index.
Since for the current graviton model there is yet no theoretically
predicted value of growth index, we attempt to provide a relevant value
 by performing a standard $\chi^{2}$
minimization procedure (described previously) between the
observationally measured
growth rate (based on the 2dF and SDSS galaxy catalogs;
see Table I; \cite{Ness08})
and that expected in the MGCDM cosmological model, according to:
\be
\chi^{2}(\gamma)=\sum_{i=1}^{5} \left[ \frac{f_{obs}(z_{i})-
f_{\rm model}(z_{i},\gamma)}
{\sigma_{i}}\right]^{2} \;\;,
\ee
where $\sigma_{i}$ is the observed growth rate uncertainty.
Note that for comparison we perform the same analysis also
for the $\Lambda$CDM model.

\begin{figure}[ht]
\mbox{\epsfxsize=8.5cm \epsffile{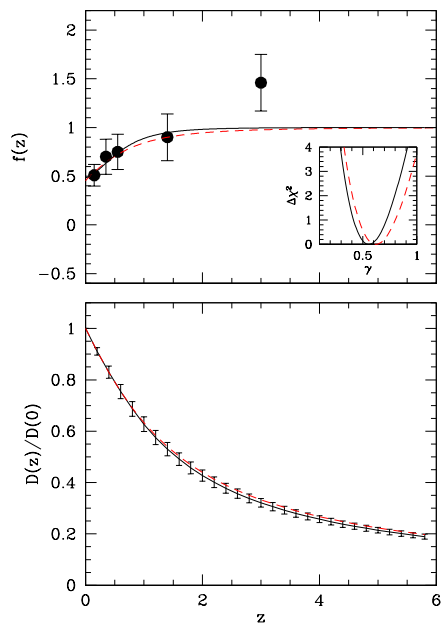}} \caption{
{\it Upper Panel:}
Comparison of the observed (solid circles\,\cite{Ness08},
(see Table I) and theoretical evolution of the growth
rate of clustering $f(z)$.
The lines correspond to the MGCDM (solid curve) and
the $\Lambda$CDM (dashed curve) models.
{\it Bottom Panel:}
The evolution of the growth factor, with that
corresponding to
the MGCDM model ($\gamma=0.56$) showing a $\sim 1-4\%$
difference with respect to that of the $\Lambda$CDM model ($\gamma_{\Lambda}=0.62$), especially at
large redshifts ($z\ge 1$). 
Errorbars are plotted only for the MGCDM model in order to avoid
confusion.}
\end{figure}

In Figure 3 (upper panel), we present
the measured $f_{obs}(z)$ (filled symbols)
with the estimated growth rate
function, $f(z)=\Omega^{\gamma}_{m}(z)$, for
the two considered cosmological models.
Notice, that for the MGCDM cosmological
model (solid line) we use $\Omega_{m}=0.276$
and for the $\Lambda$ case (dashed line) $\Omega_{m}=0.280$,  which are
the values provided by our likelihood analysis of section 3.
In the inset panel of figure 3 we plot the variation of
$\Delta \chi^{2}=\chi^{2}(\gamma)-\chi^{2}_{\rm min}(\gamma)$
around the best $\gamma$ fit value. For the MGCDM model we find
$\gamma=0.56^{+0.15}_{-0.14}$ ($\chi^{2}/dof\simeq 0.69$),
while for the $\Lambda$CDM model we obtain
$\gamma_{\Lambda}=0.62^{+0.18}_{-0.15}$ ($\chi^{2}/dof\simeq 0.75$),
which is somewhat greater, but within $1\sigma$, of
the theoretically predicted value of $\gamma_{\Lambda}\simeq 6/11$.
Such a discrepancy between the theoretical $\Lambda$CDM and observationally fitted
value of $\gamma$ has also been found
by other authors. For example, Di Porto \& Amendola
\cite{Port08} obtained $\gamma=0.60^{+0.40}_{-0.30}$, while
Nesseris \& Perivolaropoulos \cite{Ness08},
based on mass fluctuations inferred from independent observations
at different redshifts, found $\gamma=0.67^{+0.20}_{-0.17}$.
If such a systematic difference between the measured and the theoretical
$\gamma$ $\Lambda$CDM values is due to observational uncertainties or
the method used to estimate the observed $\gamma$, then one may expect
a similar systematic difference to affect the measured $\gamma$ value
for the MGCDM model, pointing to a probably more realistic value for
this model of
$\gamma\simeq 0.49$. Since however this value is within the $1\sigma$
observational uncertainty, we will consider the originally fitted
MGCDM $\gamma$ value as the nominal one.

Using the above best fit $\gamma$ values we present, in
the lower panel of figure 3, the growth factor evolution derived
by integrating Eq. (\ref{Dz221}) for the two cosmological models
(MGCDM-solid and $\Lambda$CDM-dashed). The error bars correspond to
the 1$\sigma$ uncertainty of the fitted $\gamma$ values.
Note that the growth factors are normalized to unity
at the present time. The difference between the fitted growth factors
lies, at redshifts $z\ge 1$, in the interval $\sim 1-4\%$, while
when using the theoretically predicted $\Lambda$CDM
value of $\gamma_{\Lambda}\simeq 6/11$ the difference is less than $~1.5\%$.
For a consistent treatment of the two models and for the corresponding
comparison of their respective mass functions and halo redshift distributions
we will use, throughout the rest of the paper, the
observationally derived $\gamma$ values, ie., $\gamma_{\Lambda}\simeq
0.62$ and $\gamma_{\rm MGCDM}\simeq 0.56$.

\section{\label{sec:five}Compare the cluster Halo abundances}
It is important to define observational criteria that will enable us
to distinguish between the MGCDM model and the concordance $\Lambda$CDM
cosmology. An obvious choice, that has been extensively used, is to
compare the theoretically predicted cluster-size halo
redshift distributions and to use observational cluster data to
distinguish the models. Recently, the halo abundances predicted by a
large variety of DE models have been compared with those corresponding
to the $\Lambda$CDM model \cite{Bas09c, BPL10}. As a result, such analyses
suggest that many DE models explored in this study
(including some of modified gravity)
are clearly distinguishable from the $\Lambda$CDM cosmology.

We use the Press and Schecther \cite{press} (hereafter PSc) formalism,
based on random Gaussian fields,
which determines the fraction of matter that has formed bounded
structures as a function of redshift. Mathematical details of our
treatment can be found in \cite{BPL10}; here we only present the basic
ideas.
The number density of halos, $n(M,z)$,
with masses within the range $(M, M+\delta M)$ are given by:
\begin{equation}\label{MF}
n(M,z) dM = \frac{\bar{\rho}}{M} \frac{d{\rm \ln}\sigma^{-1}}{dM} f_{\rm
PSc}(\sigma) dM \;,
\end{equation}
where $f_{\rm PSc}(\sigma)=\sqrt{2/\pi} (\delta_c/\sigma)
\exp(-\delta_c^2/2\sigma^2)$, $\delta_{c}$ is the
linearly extrapolated density
threshold above which structures collapse \cite{eke}, while
$\sigma^2(M,z)$ is the mass variance of
the smoothed linear density field, extrapolated to redshift $z$ at which
the halos are identified. It depends on the power-spectrum of
density perturbations in Fourier space, $P(k)$, for which we use here the
CDM form according to \cite{Bard86}, and the values of the
baryon density parameter, the spectral slope and Hubble constant
according to the recent WMAP7 results \cite{komatsu08}.
Although the Press-Schecther formalism
was shown to provide a good first approximation to the
halo mass function provided by numerical simulations, it was later found
to over-predict/under-predict the number of low/high mass halos at the
present epoch \cite{Jenk01,LM07}.
More recently,  a large number of works have provided better fitting
functions of $f(\sigma)$, some of them  based on a phenomenological
approach. In the present
treatment, we adopt the one proposed by  Reed et al. \cite{Reed}.

We remind the reader that it is traditional to parametrize the mass
variance in terms
of $\sigma_8$, the rms mass fluctuations on scales of $8 \;
h^{-1}$ Mpc at redshift $z=0$.

In order to compare the mass function predictions of the
different cosmological models, it is imperative to use
for each model the corresponding value of $\delta_c$ and $\sigma_8$.
It is well known that for the usual $\Lambda$ cosmology
$\delta_{c} \simeq 1.675$,
while Weinberg \& Kamionkowski \cite{Wein03} provide an accurate fitting
formula to estimate $\delta_{c}$ for any DE model
with a constant equation of state parameter.
Since for the current graviton cosmological vacuum model
the effective dark energy EoS parameter at the present
time is $w\simeq 0$ which implies that the Hubble parameter
is matter dominated it is fair to use the Einstein de-Sitter
value $\delta_{c} \simeq 1.685$ \cite{Wein03}.
Now, in order to estimate the correct model $\sigma_8$ power spectrum normalization,
we use the formulation developed in \cite{BPL10} which scales the
observationally determined $\sigma_{8, \Lambda}$ value to that of any 
cosmological model. The corresponding MGCDM value is
$\sigma_{\rm 8, MGCDM}=0.828$ and it is based on $\sigma_{8, \Lambda}=0.804$ (as
indicated also in Table 1), derived from 
an average of a variety of recent measurements (see also the corresponding discussion in 
 \cite{BPL10}) which are based on the WMAP7 results \cite{komatsu08}, on
a recent cluster
abundances analysis \cite{Rozo09}, on weak-lensing results \cite{Fu08} 
and on peculiar velocities based analyses \cite{Wat09}.

\begin{table*}[h]
\tabcolsep 10pt
\vspace {0.2cm}
\begin{tabular}{|lcc|cc|ccc|} \hline \hline
Model     & $\sigma_{8}$ & $\gamma$ &\multicolumn{2}{|c}{($\delta{\cal N}/{\cal N}_{\Lambda})_{\rm eROSITA}$} &
\multicolumn{3}{|c|}{$(\delta {\cal N}/{\cal N}_{\Lambda})_{\rm SPT}$} \\
   &  &        & $z<0.3$& $0.6\le z <0.9$ & $z<0.3$ & $0.6\le z <0.9$ & $1.3\le z <2$ \\ \hline
$\Lambda$CDM   & 0.804  & 0.62 & 0.00  & 0.00           & 0.00  & 0.00  & 0.00 \\
MGCDM          & 0.831  & 0.56 & -0.09 & 0.15$\pm 0.01$ & -0.09 & 0.11  & -0.03\\
MGCDM          & 0.875  & 0.42 &  0.00 & 0.25$\pm 0.01$ &  0.00 & 0.18  & -0.01\\
MGCDM          & 0.789  & 0.71 & -0.19 & 0.06$\pm 0.01$ & -0.19 & 0.05  & -0.05\\ \hline
\end{tabular}
\caption[]{Numerical results. The $1^{st}$ column indicates the
  cosmological model. The $2^{rd}$ and $3^{rd}$ columns lists the
corresponding $\sigma_{8}$ and $\gamma$ values, respectively.
The remaining columns present the fractional relative difference of
the abundance of halos between the
MGCDM and the $\Lambda$CDM cosmology for two future
cluster surveys discussed in the text. The lower two rows show results
corresponding to the upper and lower $1\sigma$ range of the
observational $\gamma$ value uncertainty.
Errorbars are 2$\sigma$ Poisson
uncertainties and are shown only if they are larger than $10^{-2}$).}
\end{table*}

Given the halo mass function from Eq.(\ref{MF}) we can now derive an observable
quantity which is the redshift distribution of clusters, ${\cal
N}(z)$, within some determined mass range, say $M_1\le
M/h^{-1}M_{\odot}\le M_2=10^{16}$. This can be estimated by integrating, in mass,
the expected differential halo mass function, $n(M,z)$, according
to:
\be {\cal
N}(z)=\frac{dV}{dz}\;\int_{M_{1}}^{M_{2}} n(M,z)dM, \ee
where $dV/dz$ is the comoving volume element.
In order to derive observationally relevant
cluster redshift distributions and therefore test the possibility
of discriminating between the MGCDM and the $\Lambda$CDM
cosmological models, we will use the expectations of two realistic
future cluster surveys:

\noindent (a) the {\tt eROSITA} satellite X-ray survey, with a flux
limit of: $f_{\rm lim}=3.3\times 10^{-14}$ ergs s$^{-1}$ cm$^{-2}$,
at the energy band 0.5-5 keV and covering $\sim 20000$ deg$^{2}$ of
the sky,

\noindent (b) the South Pole Telescope SZ survey, with a limiting
flux density at $\nu_0=150$ GHz of $f_{\nu_0, {\rm lim}}=5$ mJy and
a sky coverage of $\sim 4000$ deg$^{2}$.

\begin{figure}[ht]
\mbox{\epsfxsize=8.8cm \epsffile{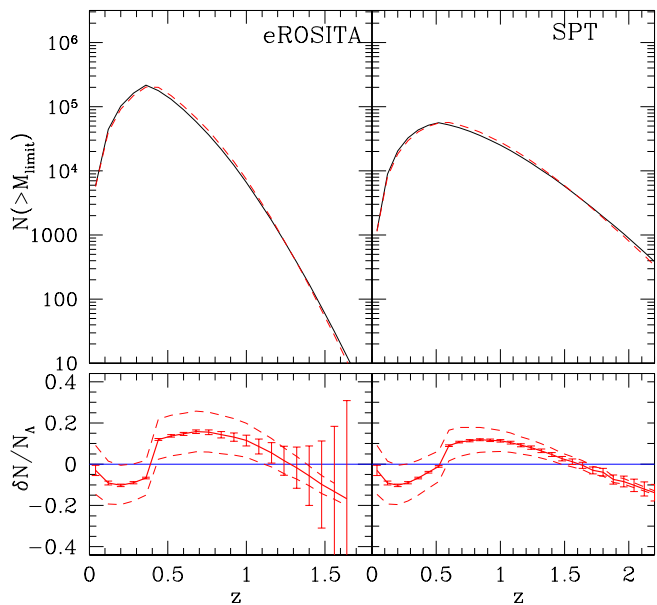}} \caption{The expected
cluster redshift distribution of the MGCDM (solid curve) and
$\Lambda$CDM (dashed curve) models
for the case of two future cluster
surveys (upper panels), and the corresponding fractional difference
with respect to the reference $\Lambda$CDM model (lower
panels). Errorbars are 2$\sigma$ Poisson uncertainties, while the
dashed lines in the lower panel bracket the range due to the uncertainty of
the observationally fitted value of $\gamma$.}
\end{figure}

To realize the predictions of the first survey we use the relation
between halo mass and bolometric X-ray luminosity, as a function of
redshift, provided in \cite{Fedeli}, ie:
\be\label{bolom}
L(M,z)=3.087 \times 10^{44} \left[\frac{M E(z)}{10^{15} h^{-1}
    M_{\odot}} \right]^{1.554} h^{-2} \; {\rm erg s^{-1}} \;.
\ee
The limiting halo mass that can be observed at redshift $z$ is
then found by inserting in the above equation the limiting
luminosity, given by: $L=4 \pi d_L^2 f_{\rm lim}${\em c}$_b$, with
$d_L$ the luminosity distance corresponding to the redshift $z$ and
{\em c}$_b$ the band correction, necessary to convert the bolometric
luminosity of Eq.(\ref{bolom}) to the 0.5-5 keV band of {\tt
eROSITA}. We estimate this correction by assuming a Raymond-Smith
(1977) plasma model with a metallicity of 0.4$Z_{\odot}$, a typical
cluster temperature of $\sim 4$ keV and a Galactic absorption column
density of $n_{H}=10^{21}$ cm$^{-2}$.

The predictions of the second survey can be realized using again the
relation between limiting flux and halo mass from \cite{Fedeli}:
\be\label{sz} f_{\nu_0, {\rm lim}}= \frac{2.592 \times 10^{8} {\rm
mJy}}{d_{A}^{2}(z)} \left(\frac{M}{10^{15} h^{-1}M_{\odot}}\right)^{1.876}
E^{2/3}(z) \; \ee where $d_A(z) \equiv d_L/(1+z)^2$ is the angular
diameter distance out to redshift $z$.

In figure 4 (upper panels) we present the expected redshift
distributions above a limiting halo mass, which is $M_1 \equiv M_{\rm
limit}=\max[10^{14} h^{-1}M_{\odot}, M_f]$, with $M_f$ corresponding to
the mass related to the flux-limit at the different redshifts,
estimated by solving Eq.(\ref{bolom}) and Eq.(\ref{sz}) for $M$. In
the lower panels we present the fractional difference between the
MGCDM and $\Lambda$CDM.
The error-bars shown correspond to 2$\sigma$ Poisson
uncertainties, which however do not include cosmic variance
and possible observational systematic uncertainties, that would
further increase the relevant variance. A further source of
uncertainty that should be taken into account is related to the
uncertainty of the observationally derived value of $\gamma$ (see
section 4). The dashed lines in the lower panels of Fig.3 bracket
the corresponding number count relative model differences due to the
$1\sigma$ uncertainty in the value of $\gamma$, with the lower curve
corresponding to $(\gamma, \sigma_8)=(0.71, 0.787)$ and the upper to
$(\gamma, \sigma_8)=(0.42, 0.876)$.

The results (see also Table II) indicate that significant model differences should be
expected to be measured up to $z\mincir 1$ for the
case of the {\tt eROSITA} X-ray survey, and to much higher redshifts
for the case of the South Pole Telescope SZ survey. What is particularly
interesting is the differential difference between the $\Lambda$CDM
and MGCDM models, which is
negative locally ($z\mincir 0.3$), positive at intermediate redshifts
($0.4\mincir z \mincir 1$) and negative again for $z\magcir 1$.
This appears to be a unique signature of the MGCDM model, which
differentiates it from the behaviour of a large class of DE models (see
\cite{BPL10}) and makes it relatively easier to be distinguished.
In Table II, one may see a more compact presentation
of our results including the
relative fractional difference between the MGCDM model and the
$\Lambda$CDM model, in characteristic redshift bins and for both
future surveys.

\section{\label{sec:six}Conclusions}
In this work,  the large and small scale dynamical properties of a flat
FLRW cold dark matter cosmology,  endowed with  massive gravitons
(MGCDM), were discussed from an analytical and a numerical viewpoints.
We find that the MGCDM can accommodate a ``dynamic phase transition"
from an early decelerating phase (driven only by cold dark matter) to a
late time accelerating expansion and a subsequent recent
re-deceleration phase.

Interestingly, the Hubble function of the MGCDM model contains
only two free parameters, namely $H_0$ and $\Omega_{m}$, which is the same
number of free parameters as the $\Lambda$CDM model.
Performing, a joint likelihood analysis using
the current observational data (SNIa, CMB shift parameter and BAOs),
we have provided
tight constraints on the main cosmological parameter
of the MGCDM model, i.e., $\Omega_{m}=0.276\pm 0.012$.
We then compared the MGCDM scenario with the conventional flat
$\Lambda$ cosmology regarding the rate of clustering as well as the
predicted halo redshift distribution.

The main conclusions of such a comparison are:
\begin{itemize}
\item At large redshifts the amplitude of the linear perturbation
growth factor of the MGCDM model is slightly different to the
$\Lambda$ solution (at a $1-4\%$ level), while the observationally determined
growth index of clustering ($\gamma\simeq 0.56$)
is smaller than the corresponding fit for
the $\Lambda$ model ($\gamma_{\Lambda}\simeq 0.62$), although within
their respective $1\sigma$ uncertainties.

\item
The shape and amplitude for the
redshift distribution of cluster-size halos predicted by the MGCDM model is quite
different from the one of a flat $\Lambda$CDM cosmology.  Such a difference depends on redshift and has a characteristic
signature that can discriminate the current graviton model from
other contender DE models in the future cluster surveys.
\end{itemize}

\vspace {0.4cm}

\acknowledgments
MP acknowledges funding by
Mexican CONACyT grant 2005-49878, and JASL is partially supported by
CNPq and FAPESP under grants 304792/2003-9 and 04/13668-0,
respectively.

\end{document}